\documentstyle[11pt,psfig]{article}
\begin{document}
\def\SUU{$SU(2)_{L}\times U(1)_{Y}$}
\def\beq{\begin{equation}}
\def\enq{\end{equation}}
\def\zz{{\cal Z}}
\def\ww{{\cal W}}
\def\ww3{{\cal W}^{3}}
\def\bb{{\cal B}}
\def\aa{{\cal A}}
\def\gblblzz{{{\overline{b_L}\gamma_{\mu} b_{L}{\cal Z}^{\mu}}\,}}
\def\gbrbrzz{{{\overline{b_R}\gamma_{\mu} b_{R}{\cal Z}^{\mu}}\,}}
\def\gblblaa{{{\overline{b_L}\gamma_{\mu} b_{L}{\cal A}^{\mu}}\,}}
\def\gbrbraa{{{\overline{b_R}\gamma_{\mu} b_{R}{\cal A}^{\mu}}\,}}
\def\gblblbb{{{\overline{b_L}\gamma_{\mu} b_{L}{\cal B}^{\mu}}\,}}
\def\gbrbrbb{{{\overline{b_R}\gamma_{\mu} b_{R}{\cal B}^{\mu}}\,}}
\def\gblblz{{{\overline{b_L}\gamma_{\mu} b_{L}{Z}^{\mu}}\,}}
\def\gblbl{{{\overline{b_L}\gamma_{\mu}\partial^{\mu} b_{L}}\,}}
\def\gbrbr{{{\overline{b_R}\gamma_{\mu}\partial^{\mu} b_{R}}\,}}
\def\zzzz{{\cal Z}_{\mu} {\cal Z}^{\mu}}
\def\wwww{{\cal W}^{+}_{\mu} {{\cal W}^{-}}^{\mu}}
\def\ttzz{{\mbox {$t$-${t}$-${\cal Z}$}\,}}
\def\bbzz{{\mbox {$b$-${b}$-${\cal Z}$}\,}}
\def\kln{\kappa_{L}^{\rm {NC}}}
\def\krn{\kappa_{R}^{\rm {NC}}}
\def\klc{\kappa_{L}^{\rm {CC}}}
\def\krc{\kappa_{R}^{\rm {CC}}}
\def\ttz{{\mbox {$t$-${t}$-$Z$}\,}}
\def\bbz{{\mbox {$b$-${b}$-$Z$}\,}}
\def\tta{{\mbox {$t$-${t}$-$A$}\,}}
\def\bba{{\mbox {$b$-${b}$-$A$}\,}}
\def\tbw{{\mbox {$t$-${b}$-$W$}\,}}
\def\tltlz{{\mbox {$t_L$-$\overline{t_L}$-$Z$}\,}}
\def\blblz{{\mbox {$b_L$-$\overline{b_L}$-$Z$}\,}}
\def\brbrz{{\mbox {$b_R$-$\overline{b_R}$-$Z$}\,}}
\def\tlblw{{\mbox {$t_L$-$\overline{b_L}$-$W$}\,}}
\def\gf{{\it G_F}}
\def\gfc{{\it G_F^{\rm 3}}}
\def\ra{\rightarrow}
\def\bar{\overline}
\def\ordera3{\mbox{${\cal O}$$(\alpha_s\gfc)$}}
\def\orderas{\mbox{${\cal O}$$(\alpha_s^3\gf)$}}
\def\Pt{\mbox{$p_T$}}
%
\def\slashchar#1{\setbox0=\hbox{$#1$}           
   \dimen0=\wd0                                 
   \setbox1=\hbox{/} \dimen1=\wd1               
   \ifdim\dimen0>\dimen1                        
      \rlap{\hbox to \dimen0{\hfil/\hfil}}      
      #1                                        
   \else                                        
      \rlap{\hbox to \dimen1{\hfil$#1$\hfil}}   
      /                                         
   \fi}                                         %
\def\centeron#1#2{{\setbox0=\hbox{#1}\setbox1=\hbox{#2}\ifdim
\wd1>\wd0\kern.5\wd1\kern-.5\wd0\fi
\copy0\kern-.5\wd0\kern-.5\wd1\copy1\ifdim\wd0>\wd1
\kern.5\wd0\kern-.5\wd1\fi}}
\def\ltap{\;\centeron{\raise.35ex\hbox{$<$}}{\lower.65ex\hbox{$\sim$}}\;}
\def\gtap{\;\centeron{\raise.35ex\hbox{$>$}}{\lower.65ex\hbox{$\sim$}}\;}
\def\gsim{\mathrel{\gtap}}
\def\lsim{\mathrel{\ltap}}
\def\D0{D\O}
\def\doublespaced{\baselineskip=\normalbaselineskip\multiply
    \baselineskip by 150\divide\baselineskip by 100}
\doublespaced
\pagenumbering{arabic}
\pagestyle{plain}
\reversemarginpar
%
\begin{titlepage}
\begin{flushright}
\today
\end{flushright}
\begin{flushright}
MSUHEP--050627 \\
CITHE--68--2006
\end{flushright}
\vspace{0.4cm}
\begin{center}
\large
{\bf High--$\bf p_T$ Higgs Boson Production
at Hadron Colliders to \ordera3}
\end{center}
\begin{center}
{\bf S. Mrenna}\footnote{mrenna@cithe502.cithep.caltech.edu}
\end{center}
\begin{center}
{Lauritsen Laboratory \\
California Institute of Technology \\
Pasadena, CA  91125}
\end{center}
\begin{center}
and
\end{center}
\begin{center}
{\bf C.--P. Yuan}\footnote{yuan@msupa.pa.msu.edu}
\end{center}
\begin{center}
{Department of Physics and Astronomy \\
Michigan State University \\
East Lansing, MI 48824}
\end{center}
\vspace{0.4cm}
\raggedbottom
\setcounter{page}{1}
\relax

\begin{abstract}
\noindent
We study high--\Pt Higgs boson
production at hadron colliders to order \ordera3~in hadron collisions.
In particular, we investigate the
process $g+q/\bar{q}\ra q/\bar{q}+H$, where $q=u,d,c,s,$ or $b$, for the LHC
(a $\sqrt{s}=$14~TeV, proton--proton collider).
Our results are compared to the \orderas~calculation.
The associated production of a high--$p_T$ Higgs boson with a $b$--quark
or anti--quark
is comparable to the \orderas~calculation because of the large top quark
mass and the additional contribution of electroweak gauge and
Goldstone bosons.  The associated production of light quarks, however,
is not significant.  We also comment on new
physics effects in the
framework of the electroweak chiral Lagrangian.
\end{abstract}

\vspace{2.0cm}
\noindent
PAC codes: 12.15.Lk, 14.80.Bn

\end{titlepage}
\newpage
\section{Introduction}
\indent

With the discovery of the top quark \cite{CDF}, the only remaining element
of the Standard Model (SM) particle spectrum is the Higgs boson.
Experimentally, there are only lower bounds on $M_H$.
LEP-I has placed the limit $M_H > 64.5$ GeV \cite{LEPH0}.
Theoretically, there are upper bounds in the SM from
unitarity and triviality arguments \cite{upperH}.
One goal of
the future High Energy Physics experimental program is to discover the
Higgs boson and verify its properties or determine the alternative
mechanism of electroweak symmetry breaking.
The search for the Higgs boson at LEP-II is strictly limited by the
available center of mass energy and luminosity,
so that only $M_H <$ 90--95 GeV can be
probed for an energy of 190 GeV and 500 pb$^{-1}$ of data \cite{LEPII}.
The reach of a high--luminosity Tevatron collider is
better, but becomes challenging above $M_H =$ 110 GeV \cite{marciano}.
The LHC, on the other hand, is hoped to have
enough energy, luminosity, and instrumentation to decisively probe the
energy scale associated with electroweak symmetry breaking.
This task is not as straight--forward as it may seem.
One possible alternative to the SM
is the Minimal Supersymmetric Standard Model (MSSM) with a constrained
multi--dimensional parameter space \cite{arnowitt,lopez,kane}.  The
constrained MSSM models predict that the couplings of the lightest Higgs
boson to SM particles is very SM--like, i.e.
$\sin^2(\beta-\alpha)\simeq 1$, and its mass should be less than about
140\,GeV \cite{susyH}.
In this case, to deduce supersymmetry,
one must observe a superpartner directly or discern its
presence in quantum corrections.  Another alternative, a model with a
strongly interacting scalar sector \cite{strongH},
predicts a greatly enhanced Higgs boson width even for a Higgs mass
of a few hundred GeV, so that
the Higgs boson signal can be hidden by backgrounds.
Regardless of the scenario, a full verification of the properties of the
Higgs boson requires a deep theoretical understanding of its properties.
In this investigation, we concentrate on the high--\Pt production of
Higgs bosons, which is sensitive to loop corrections.
The $\cal{O}$$(\alpha_s^3\gf)$ contribution to high--\Pt Higgs
boson production was calculated
previously \cite{ellis}, where $\gf = (\sqrt{2}v^2)^{-1}$ and the
vacuum expectation value $v =$ 246 GeV.
Here, we extend that calculation to include
the $\cal{O}$$(\alpha_s\gfc)$ contributions from electroweak gauge
bosons, Goldstone bosons, and quarks.  In particular, since the top
mass is large, we expect to see an enhancement in the associated production
of a Higgs boson with a $b$--quark or anti--quark in some kinematic region.
We also expect
this channel to be sensitive to the coupling of the electroweak gauge bosons
and Goldstone bosons to the Higgs boson, since it does not vanish in
the limit that the $U(1)_Y$ and $SU(2)_L$ gauge couplings
vanish.\footnote{The \orderas~contribution does not depend on
the electroweak gauge couplings, but is
only sensitive to the coupling of the top quark to the Higgs boson.}
We study this sensitivity in
the framework of the electroweak
chiral Lagrangian, which allows us to construct the
most general effective Lagrangian that is consistent with
$SU(2)_L\times U(1)_Y\ra U(1)_{em}$ symmetry breaking.
We show that with new physics, the \ordera3~contribution can be comparable
to the  \orderas~contribution for high--\Pt Higgs boson production.
Another source of high--\Pt Higgs bosons is the ${\cal O}$$(\alpha_s\gf)$
tree level process.  We show that this is small
by examining the processes $q+\bar{q}\ra b+\bar{b}+H$
and $g+g\ra b+\bar{b}+H$.
We also argue
that any interference between this order amplitude and one of higher order
is suppressed because the bottom quark mass $m_b$ is much less than the
electroweak symmetry breaking scale $v$.

\section{High--\Pt production of the Higgs boson
to \ordera3}
\indent

To \ordera3, the Higgs boson is produced at high--\Pt
from quark--gluon, antiquark--gluon, and quark--antiquark initial states.
Since they are the most interesting, we will concentrate on the first
two processes for the purpose of this discussion.  The quark--antiquark
annihilation process is typically an order of magnitude smaller at the LHC
for $M_H \le 400$ GeV.
We chose to perform
the calculation in the helicity formalism, since the amount of algebra
is reduced significantly.  Furthermore, we used the Feynman rules in the
't Hooft--Feynman gauge, since the electroweak gauge bosons and their
associated Goldstone bosons have the same mass and, hence, loop integrals
involving gauge bosons and Goldstone bosons have the same denominators.
This choice is also advantageous for investigating the electroweak
chiral Lagrangian.

We consider the process $g(p_g)+q(p_j)\ra q(p_i)+H(p_H)$ to
\ordera3,
where the $p_g$ and $p_j$ are the four-momenta of the incoming particles and
$p_i$ and $p_H$ are the four-momenta of the outgoing particles.
The quark--antiquark initial state can be generated by the substitution
$p_i\ra-p_i, p_g\ra -p_g$ and a reevaluation of the color factor.
Contributions to the loop integral come from internal lines involving
the weak isospin quark partner of $q$ (we use
the simplification $V_{ud}=V_{cs}=V_{tb}=1$),
gauge bosons, and Goldstone bosons.
Some representative Feynman diagrams are illustrated in Figure~1.
\begin{figure}
\hspace*{0.0cm}\psfig{file=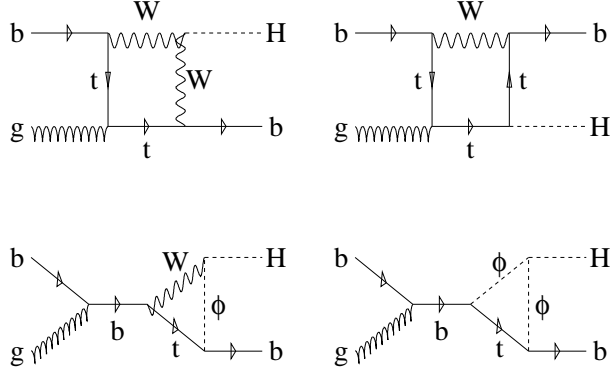,width=10cm,clip=}
\caption{Some representative Feynman diagrams
for the \ordera3~contributions to
the process $g+b/\bar{b}\ra b/\bar{b}+H$.}
\end{figure}
The amplitude for this process can
be written as:
\begin{eqnarray}
\label{eqc0}
{\cal M}_{\lambda_i \lambda_j \lambda_g} & = & i g_s
\bar{u}({\lambda_i},p_i)[{\cal F}
\gamma_{\mu}+{\cal G}_{\mu}{\slashchar{p}_{H}}]
u({\lambda_j},p_j)\epsilon^{\mu}_{\lambda_g} ,
\end{eqnarray}
where $\lambda_i$ and $\lambda_j$ are the fermion helicity indices,
$\lambda_g$ is the gluon polarization,
$\gamma_\mu$ are $4\times 4$ gamma matrices,
$\slashchar{a} = a^\mu \gamma_\mu$
for 4--vector $a^\mu$,
and equations of motion have been applied to the
on--shell, 4--component spinors.

In addition, we have taken the limit $m_q\ra 0$ for
all $q$ except $t$.  The complex scalar
${\cal F}$ and the complex vector ${\cal G}_{\mu}$ are
form factors resulting
from the integration of the loop momentum, and their explicit
expression are given in the Appendices~I and II.
The helicity amplitude can be re-written in terms of 2--component
Weyl spinors
using the bra--ket notation:\footnote{
The 2--component Weyl spinors are defined by the relation
$u_\pm={\frac{1}{2}}(1\pm\gamma_5)u$, etc.}

\begin{eqnarray*}
u_{-}(\lambda=-1/2,p_a)& = &\omega^a_{+} |p_a-\rangle, \\
v_{-}(\lambda=+1/2,p_a)& = &-\omega^a_{+} |p_a-\rangle, \\
u_{+}(\lambda=+1/2,p_a)& = &\omega^a_{+} |p_a+\rangle, \\
v_{+}(\lambda=-1/2,p_a)& = &-\omega^a_{+} |p_a+\rangle,
\end{eqnarray*}
where $\omega^a_{+}=\sqrt{2 E_a}$ for massless fermions with energy $E_a$,
and $\langle p\pm| = (|p\pm\rangle)^\dagger$.  These are the only components
for massless fermions.
For $p^\mu=(E,p s_\theta c_\phi,p s_\theta s_\phi,p c_\theta)$,
where $s_\psi$ and $c_\psi$ are
shorthand for $\sin\psi$ and $\cos\psi$,
$|p+\rangle = (\cos\theta/2,e^{i\phi}\sin\theta/2)^{\rm T}$, and
$|p-\rangle = (-e^{-i\phi}\sin\theta/2, \cos\theta/2)^{\rm T}$\footnote{
The superscript T denotes taking the transpose}.
Also, the gluon polarization 4--vectors for left-handed $(L)$ and
right-handed $(R)$ helicities can be written as:
\begin{eqnarray*}
& \epsilon^{\mu}_{\scriptscriptstyle (L)}
= \displaystyle{\frac{e^{-i\phi}}{\sqrt{2}}}
[0,i s_\phi+c_\phi c_\theta,-i c_\phi + s_\phi c_\theta,-s_\theta], & \\
& \epsilon^{\mu}_{\scriptscriptstyle (R)}
= \displaystyle{\frac{e^{i\phi}}{\sqrt{2}}}
[0,i s_\phi-c_\phi c_\theta,-i c_\phi - s_\phi c_\theta,s_\theta], &
\end{eqnarray*}
where $\phi$ and $\theta$ are spherical coordinates
of the gluon momentum.
In the helicity basis, there are four non--vanishing
amplitudes (as $m_q\ra0$).  The parton--level cross section is:
\begin{eqnarray*}
& d\hat{\sigma}
={\displaystyle\frac{1}{F}\frac{1}{S}}C_{gb}
{\displaystyle\sum_{\lambda_g=L,R}^{ }}
(|{\cal M}_{--\lambda_g}|^2 + |{\cal M}_{++\lambda_g}|^2 )dR_2,& \\
& {\cal M}_{--\lambda_g} = i g_s \omega^i_+\omega^j_+
\langle p_i -|[{\cal F}{\gamma_+}_{\mu}+{\cal G}_{\mu}
{\slashchar{p}_{H}}_+]|p_j -\rangle \epsilon^{\mu}_{\lambda_g}, & \\
& {\cal M}_{++\lambda_g} = i g_s \omega^i_+\omega^j_+
\langle p_i +|[{\cal F}{\gamma_-}_{\mu}+{\cal G}_{\mu}
{\slashchar{p}_{H}}_-]|p_j +\rangle \epsilon^{\mu}_{\lambda_g}, &
\end{eqnarray*}
where the flux factor $F = 2\hat{s}$ for
${\hat{s}}=(p_g+p_j)^2$; the spin average
factor $S = 2\times2$; the
color factor is $C_{gq}=4/(3\times 8)$ for $g+q\ra q+H$,
and $C_{qq}=4/(3\times 3)$ for $q+\bar{q}\ra g+H$;
the gluon polarization is specified by
$\lambda_g$; $dR_s$ is the two--body phase space;
$\gamma_\pm^\mu$ are the $2\times 2$ matrices
$({\bf 1},\mp\sigma_i)$,\footnote{$\sigma_i$ are Pauli matrices
satisfying Tr($\sigma_i\sigma_j$) = $2\delta_{ij}$.  }
and $\slashchar{a}_\pm = a_\mu{\gamma_\pm^\mu}$.
In the above result, ${\cal M}_{++\lambda_g}$,
which only contains contributions from $Z^0$--bosons, are small
for two reasons.  First, since we are only interested in initial and
final states without $t$--quarks, the internal quark is always
light (because of the neutral current) and
has a tiny coupling to the Goldstone boson.  Secondly, the left-- and
right--handed couplings of the $Z^0$--boson, which are smaller than the
purely left--handed coupling of the $W^\pm$--bosons, appear in the
squared matrix element to the fourth--power.  For all practical purposes,
then, the $Z^0$ contributions can be ignored, leaving only two
independent helicity amplitudes, ${\cal M}_{--\lambda_g}$,
differing only in the gluon polarization\footnote{In our numerical
results, we include all the contributions.}.

Because of gauge invariance, each amplitude satisfies the Ward Identity
resulting from replacing the gluon polarization vector with the gluon
four momentum.  This simplifies to ${\cal F}+p_g\cdot{\cal G} = 0$.
The form factors are calculated numerically using the
FF~Fortran library \cite{FF},
so the Ward Identity can be verified numerically.
Rotational and Lorentz invariance are also checked in the same manner.

The high--\Pt production of the Higgs boson at a hadron collider is
calculated by folding the parton--level cross section with the
parton distribution functions (PDF).
We use CTEQ2L parton distribution functions
and evaluate coupling constants at the momentum scale $Q^2=\hat{s}$.
Unless otherwise stated, we use $m_t$=175 GeV in all calculations.

\section{Numerical Results}
\indent

High--\Pt Higgs boson production at a 2 TeV $\bar{{\rm p}} {\rm p}$
collider is too small
to be observed for all practical purposes.\footnote{The rate at
2 TeV ($\bar{{\rm p}} {\rm p}$)
is about two orders of magnitude smaller than that at
14 TeV (${\rm p}{\rm p}$).}
We present results only for
the LHC (a 14 TeV ${\rm p}{\rm p}$ collider).
In Table~I, we list the production cross section to
\ordera3~for several Higgs boson masses as well as
the \orderas~contribution for the transverse momentum
of the Higgs boson $p_T^H>30$ GeV.
There are separate columns for the associated production
of the Higgs boson with $b$--quarks and $u,d,s,c$--quarks.  For
all of these results, we include both the quark and antiquark contribution.
For the associated production of $H$ with a $b$--quark or anti--quark, the
total \ordera3~cross section for
$p_T^H > 30$ GeV is as large as 10--20\% of the \orderas~for
$M_H = 100-200$\,GeV.
As the transverse momentum of the Higgs boson increases,
the \ordera3~contribution becomes relatively more important.
For this process, the $m_t$ dependence is minimal.
For instance, for $M_H = 110$\,GeV, the total \ordera3~cross section
of $b/\bar{b}+H$ is
16.9\,fb and 17.5\,fb for $m_t = 160$\,GeV and 190\,GeV, respectively.
For the associated production of $H$ with light quark, the
\ordera3~cross section for
$p_T^H > 30$ GeV is never more than about 3\%
of the \orderas~for $M_H = 100-200$\,GeV.
The cross section for $q+\bar{q}\ra H+g$ is much smaller than
for the corresponding process $g+q/\bar{q}\ra q/\bar{q}+H$, having values
(1.5,.15)\,fb for $M_H$ = (110,400)\,GeV, and will not be discussed
further.

We also studied the $p_T^H$ dependence of the
cross section as a function of $M_H$.
In Figure~2 we show the cross section integrated above $p_T^H$ for
$b/\bar{b}+H$ production
to \ordera3~and \orderas~in the
same $M_H$ range as in Table~I.
The mean $p_T^H$ of the \ordera3~process, for $p_T^H$ in the range
50--350 GeV, is a slowly varying function of $M_H$
below $M_H$ = 180 GeV, with a value of approximately 120 GeV.
Above $M_H$ = 180 GeV,
the mean $p_T^H$ ranges takes on the
values (139,152) GeV for $M_H$ = (200,400) GeV.
The mean $p_T^H$ of the \orderas~process is strongly dependent on the lower
transverse momentum cutoff
needed to regulate the $p_T^H \ra 0$ divergence associated with the gluon
progagator, and is somewhat smaller than that for the \ordera3~process.

\begin{table}
\renewcommand{\arraystretch}{1.33}
\begin{center}
\begin{tabular}[t]{|c|c|c|c|c|}\hline
\multicolumn{5}{|c|}{\bf Higgs Production,
$p_T^H >$ 30 GeV, $m_t$ = 175 GeV}\\
 \hline
\multicolumn{1}{|c|}{$M_H$} &
  \multicolumn{2}{|c|}{$\sigma(g+b/\bar{b} \ra b/\bar{b}+H)$ (fb)} &
  \multicolumn{2}{|c|}{$\sigma(g+q/\bar{q} \ra q/\bar{q}+H)$ (fb)}
\\ \cline{2-5}
(GeV)     & \multicolumn{1}{|c|}{${\cal O}$$(\alpha_s\gfc)$} &
       \multicolumn{1}{|c|}{${\cal O}$$(\alpha_s^3\gf)$} &
       \multicolumn{1}{|c|}{${\cal O}$$(\alpha_s\gfc)$} &
       \multicolumn{1}{|c|}{${\cal O}$$(\alpha_s^3\gf)$} \\
  \hline\hline
110  & 17.1   & 125.4 & 46.5   & 2.2$\times 10^3$   \\ \hline
140  & 15.1   & 88.8 & 37.0   & 1.6$\times 10^3$   \\ \hline
180  &  7.4   & 61.6 & 25.2   & 1.2$\times 10^3$   \\ \hline
400  &  1.9   & 29.6 &  0.5   & 0.7$\times 10^3$   \\ \hline
\hline
\end{tabular}
\end{center}
\caption{Cross section for Higgs boson production at high--\Pt.}
\end{table}

Finally, we address the issue of the ${\cal O}$$(\alpha_s\gf)$ process
$g+b/\bar{b}\ra b/\bar{b}+H$, which is a tree level process.
To estimate the size of this cross section, we used the processes
$q+\bar{q} \ra b+\bar{b}+H$ and $g+g\ra b+\bar{b}+H$
in Pythia 5.7 \cite{pythia} with
$p_T^H > 50$ GeV.  We obtained the values (4.5,3.0,2.0,.3) fb for $M_H$ =
(110,140,180,400) GeV.
In the limit that $m_b\ra0$, there is no interference between
the tree level process and the higher order
amplitudes\footnote{We note that the tree level amplitude
and the higher order
amplitudes have different helicity structure in
the $m_b\ra 0$ limit.}, so any observed
cross section is primarily the \orderas~and \ordera3~processes.

In summary, we find that to accurately predict the cross section and
test the properties of an intermediate mass Higgs boson produced at
high--$p_T$ in association with $b$ quarks and anti--quarks,
the \ordera3~contributions should be included with the \orderas~contributions.
Because of the large top quark masss, the
\ordera3~contributions are larger than the tree-level contributions of
${\cal O}(\alpha_s\gf)$ for large $p_T^H$.  Although the production rate
of \ordera3~for the associated production of the Higgs boson with a
light quark or anti--quark is not negligible, it is only a few percent of
the \orderas~rate, and therefore is probably not distinguishable from the
uncertainty in the PDF.
\begin{figure}
\hspace*{0.0cm}\psfig{file=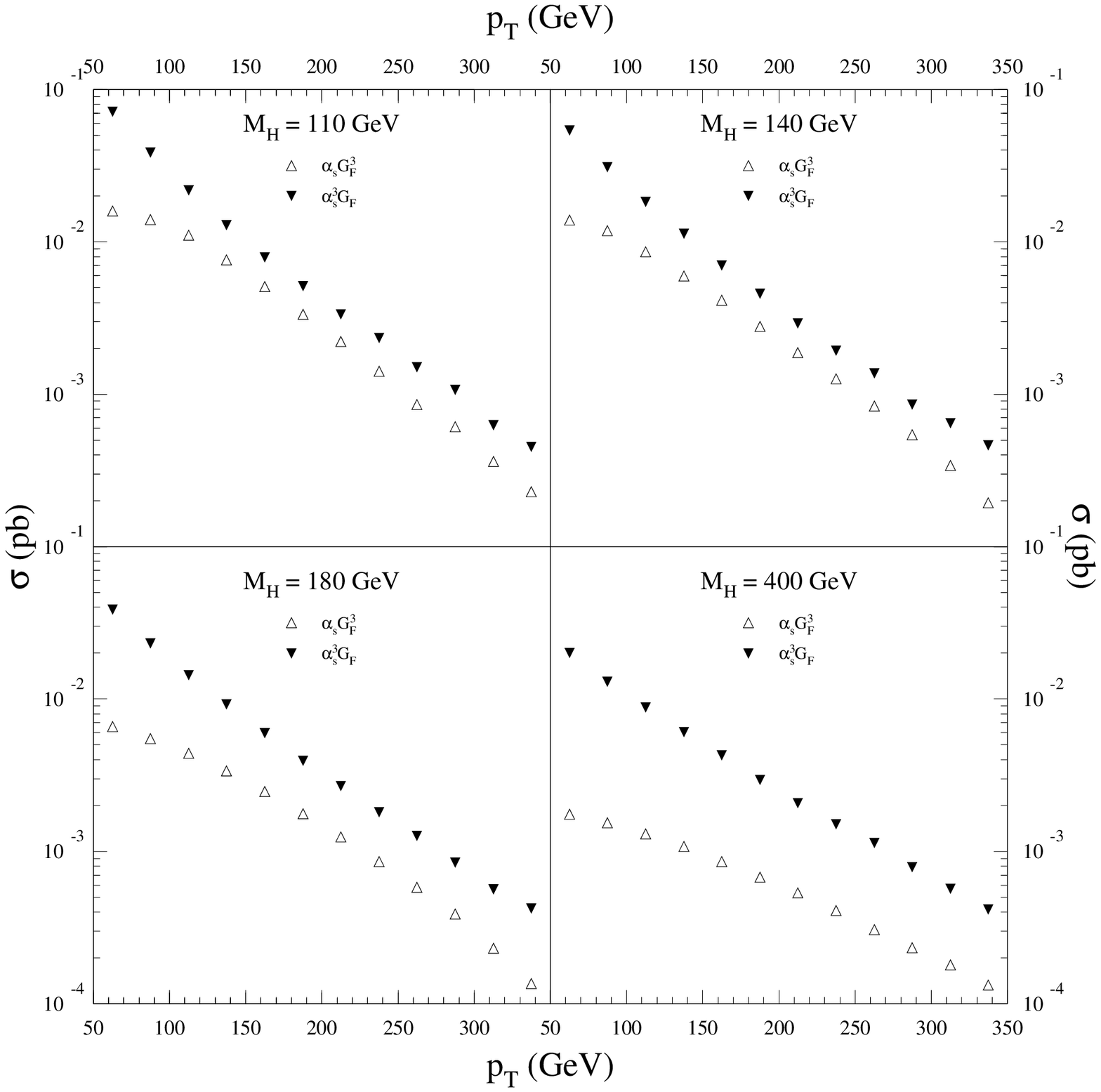,width=12cm}
\caption{The cross section integrated above $p_T^H$ for $b/\bar{b}+H$
production to \ordera3~and \orderas~for various $M_H$.}
\end{figure}

\section{The Electroweak Chiral Lagrangian and Non-Standard Model Couplings}
\indent

The process under consideration is sensitive to electroweak
symmetry breaking in three different sets of couplings.
First, there is the $t$-$b$-$W$ vertex.  The additional
non-standard couplings can be deduced from the
chiral Lagrangian \cite{tbWnew}:
\begin{eqnarray}
{\cal L}& = &-\sqrt{2}\kappa_L^{CC}\bar{t_L}\gamma^\mu{b_L}\Sigma^{+}_{\mu}
- \sqrt{2}{\kappa_L^{CC}}^\dagger\bar{b_L}\gamma^\mu{t_L}\Sigma^{-}_{\mu}
\nonumber \\
& & - \sqrt{2}\kappa_R^{CC}\bar{t_R}\gamma^\mu{b_R}\Sigma^+_\mu
 - \sqrt{2}{\kappa_R^{CC}}^\dagger\bar{b_R}\gamma^\mu{t_R}\Sigma^-_\mu,
\end{eqnarray}
where
$\Sigma^\pm_\mu = \frac{1}{\sqrt{2}} (\Sigma^1_\mu\mp i\Sigma^2_\mu)$
for $\Sigma^a_\mu = -\frac{i}{2}{\rm Tr}
(\sigma^a\Sigma^\dagger D_\mu\Sigma)$,
and the action of the covariant derivative is
$D_\mu\Sigma=\partial_\mu\Sigma-i gW^a_\mu\frac{\sigma^a}{2}\Sigma +
                                i g^{'}\Sigma B_\mu\frac{\sigma^3}{2}$.
The matrix field
$\Sigma ={\rm{exp}}\left ( i\frac{\phi^{a}\sigma^{a}}{v}\right )$,
where $\sigma^{a},\, a=1,2,3, $ are the Pauli matrices,
and $\phi^a$'s are the Goldstone bosons.
Second, there is the Yukawa coupling between the
Higgs boson and top quark.  The most general Yukawa coupling
of the fermion doublet $F$
in the chiral
Lagrangian is:
\begin{eqnarray}
{\cal L} = -\frac{c_0}{v}H \bar{F} M F,
\end{eqnarray}
where M is a $2\times 2$ mass matrix.
In the Standard Model, $c_0 = 1$.
Third, the coupling of the Higgs boson and the electroweak Goldstone
bosons comes from the Lagrangian:
\begin{eqnarray}
\label{eqc1}
{\cal L} & = &\frac{1}{2}\partial_{\mu}H\partial^{\mu} H -
{\frac{1}{2}} M_H^2 H^2 - V(H) \nonumber \\
& & + (\frac{c_1}{2} v H + \frac{c_2}{4} H^2)
{\rm Tr}\left(D_\mu\Sigma^\dagger D^\mu\Sigma\right).
\end{eqnarray}
In the Standard Model, $c_1 =c_2 = 1$.

To illustrate that new physics effects may enhance the
\ordera3~rate but not the \orderas~rate,
we study the effects of new physics arising from the scalar
sector of the Lagrangian, as shown in Eq.(\ref{eqc1}).
In this case, the \orderas~contribution is
not modified.
As shown in Ref.~\cite{strongH}, some models
of the symmetry breaking sector allow
the coefficient $c_1$ in Eq.(\ref{eqc1}) to be larger than 1.
($c_2$ is irrelevant to the processes of interest.)
For instance, $c_1=\sqrt{8/3}$ was discussed in Ref.~\cite{strongH}.
Because new physics can simutaneously modify the
interactions of $t$-$b$-$W$ and $W$-$W$-$H$, we do not intend to give
predictions for any specific model. For simplicity, we only study
the effects of new physics due to $c_1$
in the limit that the $SU(2)_L\times U(1)_Y$ gauge couplings
$g$ and $g^\prime$ vanish.
Table~II contains some of our results.

\begin{table}
\renewcommand{\arraystretch}{1.33}
\begin{center}
\begin{tabular}[t]{|c|c|c|c|c|c|}\hline
\multicolumn{6}{|c|}{\bf Higgs + $b/\bar{b}$ Rate (fb)
as $g,g^\prime\ra 0$}\\
\multicolumn{6}{|c|}{\bf
$p_T^H >$ 30 GeV, $m_t$ = 175 GeV}\\
\hline
\multicolumn{1}{|c|}{$M_H$} &
\multicolumn{5}{|c|}{Chiral Lagrangian coefficient $c_1$} \\ \cline{2-6}
(GeV) & $-.5$ & 0.0 & 0.5 & 1.0 & 1.5 \\
\hline\hline
110  & 1.09 & 0.87 & 0.78 & 0.83 & 1.01 \\ \hline
140  & 1.26 & 0.79 & 0.87 & 1.49 & 2.64 \\ \hline
180  & 0.74 & 0.69 & 2.00 & 4.67 & 8.71 \\ \hline
200  & 0.42 & 0.63 & 1.76 & 3.82 & 6.81 \\ \hline
250  & 0.19 & 0.49 & 1.32 & 2.70 & 4.60 \\ \hline
300  & 0.12 & 0.36 & 1.00 & 2.01 & 3.41 \\ \hline
400  & 0.06 & 0.21 & 0.59 & 1.20 & 2.03 \\ \hline
\hline\hline
\end{tabular}
\end{center}
\caption{Effects of new physics on high-\Pt Higgs boson
production as $g,g^\prime\ra 0$.  In the Standard Model, $c_1 = 1$.}
\end{table}

Although these rates do not represent the true rates of the process
$g+b/\bar{b} \rightarrow b/\bar{b}+H$, they illustrate that the rates can
vary by about a factor of 2 for a heavier Higgs boson.
If the electroweak corrections to high--$p_T$ Higgs production
are substantially modified by new physics at the order
${\cal O}$$(\alpha_s\gfc)$, then this can be observed in $b/\bar{b}+H$
production at future hadron colliders.

\section{Discussion and Conclusions}
\indent

Because the top quark mass is large, of the order of $v$,
the interaction of the top quark and the Goldstone bosons is
strong and, therefore, can be sensitive to the electroweak
symmetry breaking sector.
For the associated production of $b/\bar{b}$ with the Higgs boson
at high $p_T$,
the SM electroweak corrections
of \ordera3, involving the large top quark mass, is comparable to the
the QCD corrections of \orderas.
On the other hand, the associated production of light quarks and anti--quarks
with the Higgs boson is not significant because no large fermion mass is
involved.

Once the Higgs boson is discovered, it is important to test whether it is a
Standard Model Higgs boson or some
other non-standard scalar particle.
The cross section of the Higgs boson production
at large transverse momentum can be sensitive to new
physics which modify either $t$-$b$-$W$, $t$-$t$-$H$,
or $W$-$W$-$H$ vertices.  Among them, only the $t$-$t$-$H$ vertex
can modify the \orderas~contributions.
In contrast, all of them can modify the \ordera3~contributions.
As illustrated
in Table~II, it is possible that the \ordera3~rate is
enhanced by more than a factor of 2 due to new physics effects.
Therefore, \ordera3~contributions should also be included when testing SM
predictions.

\section*{ Acknowledgments }
C.-P.Y. thanks A.~Abbasabadi, D.~Bower-Chao, and W.~Repko for useful
discussions.
S.M. was supported in part by DOE grant DE--FG03--92--ER40701.
The work of C.P.Y. was supported in part by NSF grant No. PHY-9309902.

\newpage
\section*{ Appendix I: Loop Integration }
\indent

In calculating the helicity amplitudes, one must evaluate loop integrals
of the form
\begin{eqnarray*}
     X \equiv \frac{1}{i\pi^2} \int_{}^{}
     \!\!\frac{d^n Q\{1,Q^{\mu},Q^{\mu}Q^{\nu}\}}
     {(Q^2 - m_1^2)((Q+P)^2 - m_2^2)\cdots}.
\end{eqnarray*}
For triangle diagrams, $X=C$, and for box diagrams, $X=D$.

\subsection*{Triangle Diagrams}
The scalar function for triangle diagrams,
showing explicitly its dependent variables, is:
\begin{eqnarray*}
C_0(m_1^2,m_2^2,m_3^2,p_1^2, p_2^2,p_3^2) =
\frac{1}{i\pi^2} \int
     \!\!\frac{d^n Q}{(Q^2 - m_1^2)[(Q+p_1)^2 - m_2^2]
[(Q+p_1+p_2)^2 - m_3^2]},
\end{eqnarray*}
where the internal line masses $m_i$ are labelled by the external lines,
$p_1$ is the momentum flowing between the lines with masses $m_1$ and $m_2$,
$p_2$ between $m_2$ and $m_3$, and $p_3 = -p_1-p_2$ between $m_3$ and $m_1$.
The vector integral over $Q^\mu$ is
\begin{eqnarray*}
C_{11} p_1^{\mu}+C_{12} p_2^{\mu}.
\end{eqnarray*}
Similary, the tensor integral over
$Q^{\mu}Q^{\nu}$ is
\begin{eqnarray*}
C_{21}p_1^{\mu}p_1^{\nu}+C_{22}p_2^{\mu}p_2^{\nu}+
C_{23}\{p_1^{\mu}p_2^{\nu}+p_1^{\nu}p_2^{\mu}\}+C_{24}g^{\mu\nu}.
\end{eqnarray*}

\subsection*{Box Diagrams}
The scalar function for box diagrams, similar to the triangle diagrams, can
be written as
$D_0(m_1^2,m_2^2,m_3^2,m_4^2,
p_1^2,p_2^2,p_3^2,p_4^2,(p_1+p_2)^2,(p_2+p_3)^2)$.  The notation is an
obvious generalization of that for the triangle diagrams.
The vector integral over $Q^\mu$ is
\begin{eqnarray*}
D_{11} p_1^{\mu}+D_{12} p_2^{\mu}+D_{13} p_3^{\mu}.
\end{eqnarray*}
The tensor integral over
$Q^{\mu}Q^{\nu}$ is
\begin{eqnarray*}
& D_{21}p_1^{\mu}p_1^{\nu}+D_{22}p_2^{\mu}p_2^{\nu}
+D_{23}p_3^{\mu}p_3^{\nu}+& \\
& D_{24}\{p_1^{\mu}p_2^{\nu}+p_1^{\nu}p_2^{\mu}\}+
D_{25}\{p_1^{\mu}p_3^{\nu}+p_1^{\nu}p_3^{\mu}\}+
D_{26}\{p_2^{\mu}p_3^{\nu}+p_2^{\nu}p_3^{\mu}\}
+D_{27}g^{\mu\nu}. &
\end{eqnarray*}
\newpage

\section*{ Appendix II:  Form Factors}
In 't Hooft-Feynman gauge, there are 20 Feynman diagrams containing
$W^\pm$--bosons and $\phi^\pm$ Goldstone bosons involved in
the process $g+b/\bar{b}\ra b/\bar{b}+H$
at \ordera3~for $m_b = 0$.  As discussed in the text, the 3 diagrams involving
$Z^0$--bosons and $\phi^0$ Goldstone bosons are negligible.
Some typical diagrams are shown in Figure~1.
In this appendix, we list the individual contributions
to the form factors [cf. Eq.(\ref{eqc0})]
from each Feynman diagram for the
process $g(q_g)+b(q_3)\ra b(q_1)+H(q_2)$.
All momenta are defined pointing {\it in} to the Feynman diagram, i.e.
the outgoing quark $(q_1)$ and Higgs boson $(q_2)$ four momenta
have a negative energy component.  There are 12 triangle diagrams,
with terms labelled ${\cal F}_9-{\cal F}_{20}$, and 8 box diagrams,
with terms ${\cal F}_1-{\cal F}_8$ and ${\cal G}^\mu_1-{\cal G}^\mu_8$.
The full form factors are ${\cal F} = \sum_{i=1}^{20}{\cal F}_i$ and
${\cal G}^\mu = \sum_{i=1}^{8}{\cal G}^\mu_i$.
The following expressions contain the invariant masses
$s_{ij}=(q_i+q_j)^2$ and $s=(q_1+q_2+q_3)^2=q_g^2$.
We have used the relation $M_W = \frac{1}{2}g v$ to
re--express the electroweak coupling constants in terms of masses and the
vacuum expectation value $v$.
The process involving light quarks in the initial and final state
can be deduced by setting $m_t = 0$.
The limit $g,g^\prime\ra 0$, which we take to study the electroweak chiral
Lagrangian, is obtained by eliminating all terms with
an explicit $M_W$ dependence.
All triangle diagrams contain progagators for one fermion and two gauge
or Goldstone bosons.  The box diagrams fall into two categories, those
containing two fermion and two gauge or Goldstone boson propagators
(denoted $t$-$t$-$W$-$W$) and those
containing three fermion and one gauge or Goldstone boson propagators
(denoted $t$-$t$-$t$-$W$).

\subsection*{Triangle Diagrams}
\begin{eqnarray*}
      {\cal F}_9 & = & -8 C_{12} M_W^4/v^{3} \\
      {\cal F}_{10} & = & -2 C_{12} m_t^2 M_H^2/v^{3} \\
      {\cal F}_{11} & = &  2 m_t^2 M_W^2 (-2 C_{00} - C_{12})/v^{3} \\
      {\cal F}_{12} & = &  -2 m_t^2 M_W^2 (-C_{00} + C_{12})/v^{3}\\
      {\cal F}_{13}  & = &  -8 C_{12} M_W^4/v^{3} \\
      {\cal F}_{14}  & = &  -2 C_{12} m_t^2 M_H^2/v^{3}\\
      {\cal F}_{15}  & = &  2 m_t^2 M_W^2 (C_{00} - C_{12})/v^{3}\\
      {\cal F}_{16}  & = &  -2 m_t^2 M_W^2 (2 C_{00} + C_{12})/v^{3}\\
      {\cal F}_{17}  & = &  2 m_t^2 M_W^2 (-2 C_{00} - 4 C_{11})/v^{3}\\
      {\cal F}_{18}  & = &  2 m_t^4 (-C_{00} - 2 C_{11})/v^{3}\\
      {\cal F}_{19}  & = &  2 m_t^2 M_W^2 (-2 C_{00} - 4 C_{11})/v^{3}\\
      {\cal F}_{20}  & = &  2 m_t^4 (-C_{00} - 2 C_{11})/v^{3}
\end{eqnarray*}
Note that there are no tensor contributions from the triangle diagrams, i.e.
no $C_{2i}$.

\subsection*{Box Diagrams}
\begin{eqnarray*}
\mbox{$t$-$t$-$W$-$W$}~ diagrams &  \\
{\cal F}_1 & = &  -8 M_W^4 (2 D_{27} - D_{00} m_t^2 + (D_{22}-D_{24}
    + D_{25}- D_{26})M_H^2 \\
    & & + (D_{11}+D_{25})s + (D_{24}  - D_{25})s_{12}
    - (D_{11}+ D_{12}- D_{25}+ D_{26})s_{23})/v^{3} \\
{\cal F}_2 & = &
      -2 m_t^2 M_H^2 (2 D_{27} - D_{00} m_t^2 + (D_{22} - D_{24}
      + D_{25} - D_{26})M_H^2 + \\
      & & (D_{13} + D_{25})s + (D_{12} - D_{13}
      + D_{24} - D_{25})s_{12} - (D_{25} - D_{26}) s_{23})/v^{3}      \\
{\cal F}_3 & = &  -2  m_t^2 M_W^2 (-2 D_{27} + D_{13} s +
      (2 D_{00} + D_{12}- D_{13})s_{12})/v^{3}      \\
{\cal F}_4  & = &
      2  m_t^2 M_W^2 (2 D_{27} + (D_{00}+D_{11})s +(- 2 D_{00}- D_{11}
      + D_{12})s_{23})/v^{3}      \\
{\cal G}^{\mu}_1 & = &  -16 M_W^4 ((-D_{11}-
      D_{24})q_1^{\mu} +(- D_{12}- D_{22})q_2^{\mu}
      +(- D_{12} - D_{26})q_3^{\mu})/v^{3} \\
{\cal G}^{\mu}_2 & = &
      -4 m_t^2 M_H^2 ((-D_{12} - D_{24})q_1^{\mu}
      +(- D_{12} - D_{22})q_2^{\mu} +(- D_{13} -
      D_{26})q_3^{\mu})/v^{3}      \\
{\cal G}^{\mu}_3 & = &
      -4 m_t^2 M_W^2 ((-2 D_{00} - 2 D_{11} -
      D_{12} - D_{24})q_1^{\mu} \\
      & & +(- 2 D_{00} -
      3 D_{12}  - D_{22})q_2^{\mu} +(- 3 D_{13}-
      D_{26})q_3^{\mu})/v^{3}      \\
{\cal G}^{\mu}_4 & = &
      4 m_t^2 M_W^2 ((-2 D_{00}- 2 D_{11} +
      D_{12} + D_{24})q_1^{\mu}  \\
      & & +(- D_{12}+ D_{22})q_2^{\mu} +(- D_{13} + D_{26})q_3^{\mu})/v^{3} \\
\mbox{$t$-$t$-$t$-$W$}~ diagrams      \\
{\cal F}_5  & = &  4 m_t^2 M_W^2 (D_{00} m_t^2 +(- D_{22} + D_{24}
      - D_{25} + D_{26})M_H^2  - (2 D_{11}+ D_{25}) s \\
   & &+(- D_{24}+ D_{25})s_{12}
   +(- D_{00}+ 2 D_{11} - 2 D_{12} + D_{25} - D_{26})s_{23})/v^{3} \\
{\cal F}_6  & = & 2 m_t^4 (D_{00} m_t^2 +
     (- D_{22} + D_{24} - D_{25} + D_{26})M_H^2 - (2 D_{13} s + D_{25}) s \\
   & & +(- D_{00} - 2 D_{12} + 2 D_{13} - D_{24} + D_{25})s_{12}
   + (D_{25} - D_{26})s_{23})/v^{3}      \\
{\cal F}_7  & = &   4 m_t^2 M_W^2 (D_{00} m_t^2 +
      (- D_{22}+ D_{24}- D_{25}+ D_{26})M_H^2 \\
      & & + (D_{00}- D_{11}+ D_{13}- D_{25})s +(- D_{00}
      + D_{12}- D_{13}- D_{24}+ D_{25})s_{12} \\
      & & + (D_{11}- D_{12}+ D_{25}  - D_{26})s_{23})/v^{3} \\
{\cal F}_8  & = &  2 m_t^4 (D_{00} m_t^2 +(-D_{22}+ D_{24}- D_{25}
      + D_{26})M_H^2 \\ & & + (D_{00}+ D_{11}- D_{13}- D_{25})s
      +(- D_{12}+ D_{13}- D_{24} + D_{25})s_{12} \\
      & &+(- D_{00}- D_{11}+ D_{12}
      + D_{25}- D_{26})s_{23})/v^{3}      \\
{\cal G}^{\mu}_5 & = &
      8 m_t^2 M_W^2 ((3 D_{11} +
      2 D_{24})q_1^{\mu} + (D_{00} + 3 D_{12} +
      2 D_{22})q_2^{\mu} \\
      & & + (D_{00} + 2 D_{12} +
      D_{13} + 2 D_{26})q_3^{\mu})/v^{3}       \\
{\cal G}^{\mu}_6 & = &  4 m_t^4 ((D_{00}+ D_{11}+
      2 D_{12}+ 2 D_{24})q_1^{\mu} + (D_{00} +
      3 D_{12} + 2 D_{22})q_2^{\mu} \\
      & & + (3 D_{13} + 2 D_{26})q_3^{\mu})/v^{3}      \\
{\cal G}^{\mu}_7  & = &  8 m_t^2 M_W^2 ( (D_{11} +
      2 D_{24})q_1^{\mu} + (D_{12} + 2 D_{22})q_2^{\mu} \\
      & & +(- D_{00} + 2 D_{12}- D_{13}+
      2 D_{26})q_3^{\mu})/v^{3}      \\
{\cal G}^{\mu}_8 & = &  4 m_t^4 ((-D_{00} - D_{11}+
      2 D_{12} + 2 D_{24})q_1^{\mu} + (D_{12}  +
      2 D_{22})q_2^{\mu} \\
      & & + (D_{13} + 2 D_{26})q_3^{\mu})/v^{3}
\end{eqnarray*}

\end{document}